\documentclass[preprint,3p,sort&compress]{elsarticle}
\usepackage{amsmath}	
\usepackage{amssymb}
\usepackage{color}
\usepackage{graphicx}
\usepackage{slashed}
\makeatletter
\def\ps@pprintTitle{%
 \let\@oddhead\@empty
 \let\@evenhead\@empty
 \def\@oddfoot{\centerline{\thepage}}%
 \let\@evenfoot\@oddfoot}
\makeatother

\def\QCD{\ensuremath{\mathrm{QCD}}}
\def\QED{\ensuremath{\mathrm{QED}}}
\def\as{\ensuremath{\alpha_s}}
\def\asQCD{\ensuremath{\as'}}
\def\asSM{\ensuremath{\as}}
\def\asQCDbare{\ensuremath{\as'{}_0}}
\def\asSMbare{\ensuremath{\as{}_0}}
\def\al{\ensuremath{\alpha}}
\def\abare{\ensuremath{\alpha{}_0}}
\def\a{\ensuremath{\alpha}}
\def\MSbar{\ensuremath{\overline{\mathrm{MS}}}}

\def\sw{\ensuremath{s_W}}
\def\cw{\ensuremath{c_W}}

\def\dscale{\ensuremath{\mu}}

\def\Mhard{\ensuremath{M}}
\def\Mhardbare{\ensuremath{M_0}}

\def\mtpole{\ensuremath{m_t}}
\def\mtrun{\ensuremath{\hat m_t}}

\def\mmt{\ensuremath{\mtpole^2}}
\def\mmb{\ensuremath{m_b^2}}
\def\mmW{\ensuremath{m_W^2}}
\def\mmZ{\ensuremath{m_Z^2}}
\def\mmh{\ensuremath{m_h^2}}

\def\xwt{\ensuremath{x_{wt}}} 
\def\xwz{\ensuremath{x_{wz}}} 
\def\xzt{\ensuremath{x_{zt}}}
\def\xht{\ensuremath{x_{ht}}}

\def\ytsC{\ensuremath{\left( \frac{\mmt}{2 \mmW \sw^2}\right)}}
\def\ybsC{\ensuremath{\left( \frac{\mmb}{2 \mmW \sw^2}\right)}}
\newcommand{\dzeta}[2]{ \ensuremath{ \delta \zeta_{#1}^{(#2)} } }
\newcommand{\dbeta}[2]{ \ensuremath{ \beta_{#1}^{(#2)} } }
\newcommand{\XX}[2]{ \ensuremath{X^{(#1)}_{#2}}}
\def\lnMTmu{\ensuremath{\ln \frac{\mmt}{\dscale^2}}}
\def\lnSqMTmu{\ensuremath{\ln^2 \frac{\mmt}{\dscale^2}}}

\journal{Physics Letters B}
\begin{document}
\begin{frontmatter}

\title{On the electroweak contribution to the matching of the strong coupling constant in the SM}
\author[a]{A.~V.~Bednyakov}
\ead{bednya@theor.jinr.ru}

\address[a]{Joint Institute for Nuclear Research,\\
 141980 Dubna, Russia}

\begin{abstract}
The effective renormalizable theory describing electromagnetic and strong interactions of quarks of five light flavors 
($n_f = 5$ $\QCD\times\QED$) is considered as a low-energy limit of the full Standard Model. 
Two-loop relation between the running strong coupling constants $\as$ defined in either theories is found by
simultaneous decoupling of electroweak gauge and Higgs bosons in addition to the top quark.
The relation potentially allows one to confront ``low-energy'' determination of $\as$ with a high-energy one
with increased accuracy. 
Numerical impact of new $\mathcal{O}(\as\al)$ terms is studied at the $M_Z$ scale.
It is shown that the corresponding contribution, although being suppressed with respect to $\mathcal{O}(\as^2)$ terms, 
is an order of magnitude larger than the three-loop QCD corrections $\mathcal{O}(\as^3)$ usually taken into account
in four-loop renormalization group evolution of $\as$. The dependence on the matching scale 
is also analyzed numerically.

\end{abstract}

\end{frontmatter}

The strong coupling constant $\as$ being a fundamental parameter of the Standard Model (SM) 
	Lagrangian is not predicted by 
	the model and should be determined from experiment.
The theory of strong interactions, Quantum Chromodynamics (QCD), embedded in the SM,  should allow one to relate 
different observables measured at different scales. 
In perturbation theory (PT) one usually employs the notion of running coupling $\as(Q^2)$ (see Ref.~\cite{Moch:2014tta} for a recent discussion on its determination) 
	depending on some characteristic scale $Q$ of the considered process,
	so that predictions are typically given by (truncated) series in $\as(Q^2)$.
The dependence of $\as$ on $Q^2$ is given through the renormalization group (RG) equation
\begin{equation}
	\frac{d \as(Q^2)}{d \ln Q^2} = \beta ( \as(Q^2) ). 
	\label{eq:as_rge_general}
\end{equation}
A proper choice of $Q^2$ allows one to sum up a certain type of logarithmic corrections appearing at each 
	order of PT.
In the minimal $\MSbar$ renormalization scheme~\cite{'tHooft:1973mm} beta-functions have a simple polynomial form and are known up to the 
	four-loop level \cite{vanRitbergen:1997va,Czakon:2004bu}.
However, it is a well-known fact (see, e.g., the  pioneering work \cite{Bernreuther:1981sg} and reviews in Refs.~\cite{Steinhauser:2002rq,Grozin:2012ec}) 
that in the models with very different mass scales $m\ll M$ 
	one needs to employ the ``running-and-decoupling'' procedure to re-sum large logarithms involving ratios of particle masses $\log m/M$
	in the ``low-energy'' observables.\footnote{Related to processes with characteristic scale $Q \lesssim m$. }
Application of this technique to perturbative calculations results in the absorption of leading effects due to heavy degrees of freedom with mass $\mathcal{O}(M)$ 
	in the parameters of the effective theory (ET). 
	This procedure is usually applied in QCD \cite{Chetyrkin:1997un,Chetyrkin:2005ia,Schroder:2005hy,Chetyrkin:2005ia,Kniehl:2006bg,Grozin:2011nk}
	to decouple (``integrate out'') heavy quarks and define running $\as^{(n_f)}(\mu)$ in the effective $n_f$ - flavor theory.\footnote{Quark running masses are also affected by decoupling (see Ref.~\cite{Chetyrkin:2000yt}
	and references therein).}
In addition to this, a study of matching corrections~\cite{Jegerlehner:2012kn,Kniehl:2014yia,Degrassi:2012ry} is unavoidable if one is interested in high-energy behavior of the SM (see., e.g.,~\cite{Bezrukov:2012sa,Degrassi:2012ry}).
The latter is analyzed with the help of the RGEs~\cite{Luo:2002ey,Mihaila:2012fm,Chetyrkin:2012rz,Bednyakov:2012rb,Chetyrkin:2013wya,Bednyakov:2013eba,Bednyakov:2013cpa,Bednyakov:2014pia}
	which take into account all the interactions of the SM.

It is also worth mentioning that the same method can be applied to the supersymmetric (SUSY) extensions of the SM: 
	SUSY-QCD corrections are considered in Refs.~\cite{Harlander:2005wm,Bednyakov:2007vm,Bauer:2008bj,Kurz:2012ff}
	and leading corrections due to electroweak interactions are calculated in Refs.~\cite{Bednyakov:2009wt,Noth:2010jy}.

The value of the running strong coupling $\as(Q)$ can be determined 
	(see Ref.~\cite{Agashe:2014kda} for a comprehensive review and references therein) 
	from a bunch of experiments with a characteristic scale $Q$ ranging 
	from the tau-lepton mass $m_\tau=1.77682(16)$ GeV up to about 1 TeV.
In addition, electroweak precision fits and 
	lattice QCD calculations can be used to 
	yield a value for $\as$. 
In order to compare different measurements and determinations with each other and use them in a global analysis of QCD,  effective couplings extracted from experiments
	are converted to the \MSbar-scheme and evolved by means of RGEs \eqref{eq:as_rge_general} to the reference scale which is chosen to be $M_Z$.

In the above-mentioned RGE analysis one usually neglects the influence of electroweak interactions on the strong coupling. 
However,  it is obvious that since QCD is embedded in the SM, virtual electroweak bosons can also modify the strength of the strong interactions (and vice versa).
This effect, however, starts at the two-loop level.
The aim of the present paper is to apply the decoupling procedure to find a two-loop relation between the strong coupling defined in the $n_f=5$ $\QCD \times \QED$ effective theory  
	and that of the full SM. 
Additional terms due to integrated $W$-, $Z$-, and Higgs bosons can be used to decrease an uncertainty in the prediction of the SM strong coupling $\as(M_Z)$
	from, e.g., $\as(m_\tau)$  extracted from tau-lepton decays, or, vice versa, a more accurate estimate of the effective $\as$ can be made given the SM input.

Let us give a brief description of the calculation techniques together with the approximations employed.
It is convenient to carry out matching at the level of Green functions\footnote{One can also consider observables for matching.} 
of light fields by comparing the results obtained in the effective and a more fundamental theory.
By integrating out heavy degrees of freedom
(the top-quark with mass $\mmt$, the $W$- and $Z$- bosons with masses $\mmW$ and $\mmZ$, respectively, and the Higgs boson with mass $\mmh$) 
	we obtain an effective (``low-energy'') description of the SM 
	valid far below the electroweak scale. 
The latter is parametrized by an effective Lagrangian, which involves non-renormalizable operators 
	in addition to the renormalizable ones.
Both types of operators contribute to the Green functions of ET. 
	However, the latter are suppressed by the inverse power of some large mass scale.  
The couplings for the ET operators can be deduced from the parameters of a more fundamental theory, the SM in our case,
	by means of matching (decoupling) procedure.

In theories with spontaneously broken symmetries the application of the decoupling procedure 
	suffers from the following subtlety. 
Since particle masses are related to the corresponding Higgs couplings, a large mass limit can be obtained
	either by setting the Higgs field vacuum expectation value (VEV) $v$ to infinity or by assuming that
	$v$ is fixed but (some of) the couplings (e.g., the top quark Yukawa coupling) tend to infinity
	(see, e.g., \cite{Pich:1998xt} and references therein).
In literature, one usually utilizes the latter option 
	and speaks of ``non-decoupling'' feature of the top quark since
	the corresponding Yukawa coupling is expressed in terms of its mass. 

In this paper, we assume that the decoupling limit is obtained by setting $v\to\infty$, 
so that non-renormalizable Fermi-type operators are neglected. 
Nevertheless, a certain hierarchy in the Higgs couplings is assumed.
Light quarks, which are not integrated out and are ``left'' in ET, 
	are considered to be massless and, as a consequence, have vanishing Yukawa couplings to the Higgs boson.  
All other Higgs couplings are treated on equal footing. 

An additional comment regarding the neglected interactions of the Higgs boson is in order. 
If we want to take, e.g., $b$-quark Yukawa coupling into account, %in the decoupling of $\alpha_s$, 
	we inevitably have to consider the matching of non-renormalizable operators in ET (e.g., Fermi-type operators).
This is due to the fact that both the dimensionless Yukawa couplings in the full theory and the non-renormalizable ET interactions, which are formally suppressed by $\mmW$, can lead 
	to comparable contributions $\mathcal{O}(\mmb/\mmW)$ to the Green functions utilized for matching.
Our setup allows us to circumvent this difficulty and avoid matching of non-renormalizable ones,
	which are suppressed by the ratio of ``soft'' (in our setup it is either some external momentum 
	or the mass of a light quark) and ``hard'' (electroweak) scale.  

In the considered problem electroweak interactions can only appear in loops
	involving quarks so %it is obvious 
	that the electroweak bosons contribute starting with the two-loop level.  
Due to this, the relation between the strong coupling constants defined in the effective five-flavor $\QCD\times\QED$ (denoted by $\asQCD$) and  the full SM ($\asSM$)
has the following form:
\begin{eqnarray}
		\asQCD  & = &  \asSM \zeta_{\as} = \asSM \left( 1 
		+ \frac{\asSM}{4\pi}	\dzeta{\as}{1} 
		+  \frac{\asSM^2}{(4\pi)^2}~\dzeta{\as}{2} 
		+ \frac{\asSM \al}{(4\pi)^2}~\dzeta{\as\al}{2} 
		+ \ldots
		\right),
\label{eq:as_dec}
\end{eqnarray}
	where the running couplings are assumed to be renormalized in the $\MSbar$  scheme and the dependence on the decoupling scale $\mu$ is implied. 
The strong coupling $\asSM$ and the fine-structure constant $\al$ in the right-hand side (RHS) of Eq.~\eqref{eq:as_dec} are defined in the full SM.
Pure QCD corrections to the decoupling constant $\zeta_{\as}$ were calculated quite a long time ago \cite{Larin:1994va} 
	and are given at the two-loop level by the expressions ($C_A=3$, $C_F = 4/3$, $T_F=1/2$):
\begin{eqnarray}
% 1 loop
	\dzeta{\as}{1} & = & X^{(1)}_{\as} \lnMTmu , \qquad X^{(1)}_{\as} = \frac{4}{3} T_f = \frac{2}{3} %\ln \frac{\mmt}{\dscale^2},
\label{eq:as_dec_as_contrib}
		\\
% 2loop
		\dzeta{\as}{2} & = & 
		X^{(0)}_{\as^2} + X^{(1)}_{\as^2} \lnMTmu + X^{(2)}_{\as^2} \lnSqMTmu 
\label{eq:as_dec_asas_contrib} 
\\
X^{(0)}_{\as^2} & = & \left( \frac{32}{9} C_A - 15 C_F \right) T_f =  -\frac{14}{3} 
\nonumber\\
X^{(2)}_{\as^2}  & = &   \frac{16}{9} T_f^2 = \frac{4}{9},\qquad
X^{(1)}_{\as^2}  =  \left( \frac{20}{3} C_A + 4 C_F \right) T_f  = \frac{38}{3}
\end{eqnarray}
 	in which $\mtpole$ corresponds to the top quark pole mass. 
The latter can be expressed in terms of the running mass $\mtrun$  in perturbation theory.
However, the advantage of $\mtpole$ lies in the fact that it corresponds  to an ``observable'' 
(modulo subtleties mentioned in Refs.\cite{Fleming:2007qr,Moch:2014tta}) quantity.
In addition, this choice allows one to keep all the dependence of $\dzeta{\as}{i}$ on the matching scale $\mu$ explicit.
Since we are interested in the electroweak corrections, the relation between $\mtrun$ and $\mtpole$
	should be considered in the full SM. 
Let us mention a crucial role of tadpole diagrams
rendering the corresponding running mass $\mtrun$ a gauge-independent quantity (see, e.g., Ref.\cite{Fleischer:1980ub}  and references therein).
Initially, the result for $\dzeta{\as\al}{2}$ has been obtained in terms of running parameters in the $\MSbar$ scheme (with the account 
of tadpole diagrams as in Ref.~\cite{Awramik:2002vu}) and latter recalculated with the on-shell counter-term for the top-quark mass.
As expected, the tadpole contribution to the considered quantity 
	was canceled by the counter-term allowing one from the very beginning to ignore the tadpole issue.
The price to pay for this kind of simplifications is gauge-dependence of the top quark mass counter-term.
Having this in mind, in what follows we present the expression for $\dzeta{\as\al}{2}$ in terms of ``physical'' masses 
referring to the well-known one-loop relation between the pole and running quark masses \cite{Hempfling:1994ar,Jegerlehner:2002em,Jegerlehner:2012kn}  
	in the Standard Model.\footnote{For consistency one should neglect all masses but $\mmt$, $\mmW$, $\mmZ$ and $\mmh$ 
		when expressing $\mtpole$ in terms of $\mtrun$.}

For the calculation of $\dzeta{\as\al}{2}$ %of strong coupling decoupling constant 
we have used the fact that the renormalized decoupling constant can be deduced 
from the relation between the bare couplings\footnote{Considered in dimensionally regularized theory with space-time dimension $d=4-2\varepsilon$.} \cite{Chetyrkin:1997un}, denoted by $\asQCDbare$ and $\asSMbare$, after proper renormalization, i.e., 
\begin{equation}
	\asQCDbare = \zeta_{\as,0}\left[\abare, \Mhardbare \right] \times \asSMbare \Rightarrow 
    \asQCD = \frac{Z_{\as}\left[\a\right]}
    		      {Z'_{\as}\left[\asQCD\right]} 
                  \times
    	\zeta_{\as,0}\left[Z_{\a} \a, Z_{\Mhard} \Mhard \right] \times \asSM.
\label{eq:dec_bare_to_ren}
\end{equation}
	Here $\zeta_{\asSMbare}$ is a bare decoupling constant,
    $\Mhard$ corresponds to the masses of heavy particles 
    that should be "integrated out", and $\a$ in the RHS collectively denotes all couplings of the SM.

The required renormalization constants in $\QCD\times\QED$ and the SM can be written in the following form (up to the two-loop order):
\begin{eqnarray}
	Z_{\as} & = & 
     1  + \frac{\as}{4\pi} \frac{\beta^{(1)}_{\as}}{\varepsilon}
		+ \frac{\as{}^2}{(4\pi)^2}
        \left[
       \left(\frac{\dbeta{\as}{1}}{\varepsilon} \right)^2 
		+ \frac{\dbeta{\as}{2}}{2\varepsilon}
		\right]
        +
        	\frac{\as \al}{(4\pi)^2}
            \frac{\dbeta{\as\al}{2}}{2\varepsilon}
		\label{eq:asMS_ren}
\end{eqnarray}
where in $\QCD\times\QED$ we have (see, e.g., Ref.~\cite{Surguladze:1996hx})
\begin{eqnarray} 
    &&\dbeta{\asQCD}{1}  =  \dbeta{\as}{1}(n_f = 5),\qquad \dbeta{\asQCD}{2}  =  \dbeta{\as}{2}(n_f = 5), 
    \label{eq:beta_as_et} 
    \\
    &&\dbeta{\asQCD\al}{2} =  4 T_F \left[ n_u Q_u^2 + n_d Q_d^2 \right], \qquad n_u = 2,~n_d = 3, 
    \qquad Q_u = \frac{2}{3},~Q_d = -\frac{1}{3},
\end{eqnarray}
	in which $n_u(n_d)$ counts the number of active $u$($d$)-quarks with charges $Q_u$($Q_d$) in ET.

In the SM (see Refs.~\cite{Luo:2002ey,Mihaila:2012fm,Bednyakov:2012rb}) the corresponding expressions 
	look like 
\begin{eqnarray}
    &&\dbeta{\as}{1}  =  \dbeta{\as}{1}(n_f = 6),\qquad \dbeta{\as}{2}  =  \dbeta{\as}{2}(n_f = 6), 
    \label{eq:beta_as_sm} 
    \\
    &&\dbeta{\as\al}{2} =  4 T_F \left[ - \ytsC  - \ybsC  + \frac{11}{12\cw^2}+\frac{9}{4\sw^2} \right].
    \label{eq:as_asal_sm}
\end{eqnarray}
In Eqs.~\eqref{eq:beta_as_et}-\eqref{eq:beta_as_sm} the pure QCD beta-function is given by~\cite{Gross:1973id,Politzer:1973fx,Caswell:1974gg,Jones:1974mm,Tarasov:1976ef,Egorian:1978zx}
\begin{eqnarray}
		\dbeta{\as}{1}(n_f) & = & - \frac{11}{3} C_A + \frac{4}{3} T_f n_f,
	\label{eq:beta_qcd_ol} \\
		\dbeta{\as}{2}(n_f) & = & 
		- \frac{34}{3} C_A^2 + \frac{20}{3} C_A T_f n_f + 4 C_F T_f n_f .
	\label{eq:beta_qcd_tl}
\end{eqnarray}
	In \eqref{eq:as_asal_sm} both quark Yukawa, $y_t$ and $y_b$, 
	and fundamental electroweak SU(2)$\times$U(1) gauge couplings $g_2$, $g_1$
	are rewritten in terms of running particle masses and the fine-structure constant,
	which is factorized, i.e. the following relations are assumed to be valid for the running 
	\MSbar-parameters
\begin{eqnarray}
	e = g_1 \cw = g_2 \sw,\qquad \cw^2 = \frac{g_2^2}{g^2_2+ g_1^2} = \frac{\mmW}{\mmZ}, \qquad y_q = \frac{e m_q}{\sqrt 2 m_W \sw}. 
\label{eq:run_ps_rel}
\end{eqnarray}
	As it was mentioned earlier, in what follows we will neglect 
	the $b$-quark mass, i.e., we set $y_b = m_b = 0$.

\begin{figure}[th]
\includegraphics[width=\textwidth]{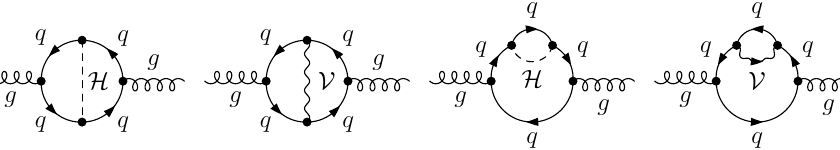}
\caption{
\label{fig:diagrams}
  Gluon self-energy diagrams giving rise to the $\mathcal{O}(\alpha_s \alpha)$ contribution to the strong coupling decoupling constant. The fields denoted by $q$ correspond to different quarks, $\mathcal{H}=h, G_0, G^\pm$ are the SM Higgs and would-be goldstone bosons. The heavy electroweak gauge bosons are given by $\mathcal{V}=W^\pm,Z$.}
\end{figure}

The bare decoupling constant for the strong coupling is obtained by considering ghost-ghost-gluon vertex.\footnote{
It is worth pointing that the same result has also been obtained by considering the $b$-quark - gluon vertex.} 
    Given the corresponding bare decoupling constant - $\zeta_{cGc}$, and that of gluon ($\zeta_{G}$) and ghost ($\zeta_{c}$) propagators, one  
    can write (omitting the "bare" labels)
\begin{equation}
\zeta_{\alpha_s} =  \zeta^2_{cGc} \, \zeta_{c}^{-2} \, \zeta_{G}^{-1}.
\label{eq:bare_as_cGc}
\end{equation}
	Here $\zeta_{cGc},~\zeta_{c}$ and $\zeta_{G}$ are
    determined by the leading coefficient in Taylor expansion 
    of the corresponding bare Green functions in small external momenta and masses.  
Since Feynman integrals with no mass scale vanish in dimensional regularization, only
      vacuum integrals with at least one massive line survive.
Due to the fact that gluons and ghosts do not couple directly to the Higgs and electroweak bosons,
	all the relevant diagrams giving rise to the new two-loop contribution 
	$\dzeta{\as\al}{2}$ are presented in Fig.~\ref{fig:diagrams}.
	The corresponding Feynman amplitudes were generated by means of the {\tt FeynArts} package~\cite{Hahn:2000kx}.
Two-loop vacuum integrals, which appear after the Taylor expansion, were recursively reduced to a master integral~\cite{Davydychev:1992mt}
	via the integration-by-parts method~\cite{Chetyrkin:1981qh}.

Since both the renormalization and bare decoupling constants contain poles in 
$\varepsilon$, the finiteness of the final result for $\zeta_{\as}$ 
	provides a useful test of the valididy of calculation.
In addition, the cancellation of electroweak gauge-fixing parameters in 
	the RHS of \eqref{eq:as_dec} allows one to cross check the obtained result.

After proper  renormalization the final expression for the electroweak contribution 
is given by %the expression 
\begin{eqnarray}
%% 2loop ew
		\dzeta{\as\al}{2} & = & \frac{\mmt}{\mmW \sw^2} 
		\left( X^{(1)}_{\as\al} \lnMTmu + X^{(0)}_{\as\al}  \right).
	\label{eq:as_dec_asal_contrib_} 
\end{eqnarray}
For convenience, in expression \eqref{eq:as_dec_asal_contrib_} we have factored out a large 
	ratio\footnote{On-shell value for 
	$\sw^2 = 1 - \mmW/\mmZ$ is used.} 
$\mmt/(\mmW\sw^2) = 20.8(2)$, which corresponds to the dominant top-quark Yukawa coupling contribution. 

The coefficients $\XX{0,1}{\as\a}$ can be written in 
the following form ($x_{ij} \equiv m_i/m_j$, e.g., $\xwt = m_W/\mtpole$):
{\allowdisplaybreaks
\begin{eqnarray}
		X^{(1)}_{\as\al} & = & -1 + \underbrace{\xwt^2 \left( \frac{2}{9} %\Xwt 
			+ \frac{22}{9} \xwz^2 \right) %\frac{Xwt^2}{\Xzt} 
			+ \frac{11}{6} \xzt^2}_{0.955446}, %\Xzt
	\label{eq:as_dec_asal_log_contrib} 
			\\
		X^{(0)}_{\as\al} & = & 
%% 1
	\left(1-\frac{\xzt^2}{4}\right)^{-1} \left[ 
	\underbrace{\frac{5}{4}}_{1.34306}	
	+ \underbrace{\left(-\frac{1}{2} \xwt^2 \xwz^2-\frac{5 \xwt^2}{6}-\frac{11 \xzt^2}{6}\right)}_{-0.82871} 
		\right.
	\nonumber\\
	&-&
		 \underbrace{\left(\frac{197 \xwt^4}{216}-\frac{235 \xwt^2 \xzt^2}{216}-\frac{19 \xzt^4}{864}\right)}_{0.0261372}
	\left.
		+ \underbrace{\left(\frac{25 \xwt^4 \xzt^2}{108}-\frac{5 \xwt^2 \xzt^4}{27}+\frac{17 \xzt^6}{216}\right)}_{0.00170613}
\right]
	\nonumber\\
% 2 + 3
	&-&
\underbrace{\frac{4}{3} \xht \left(1-\frac{\xht^2}{4}\right)^{3/2} \arccos\left(\frac{\xht}{2}\right)}_{-0.939121}
\nonumber\\
& +&  \frac{\xwt^2}{1-\xwt^2}%^{-1} 
	\log \left(\xwt^2\right) %\times 
	\times\left(
	\underbrace{\vphantom{\frac{1}{2}}2}_{-0.842913} 
	-\underbrace{\frac{7  \xwt^2}{2}}_{0.317706} 
	+\underbrace{\frac{5  \xwt^4}{6}}_{-0.0162922}
	-\underbrace{\frac{1}{3}  \xwt^6}_{0.0014036}
	\right) 
	\nonumber \\
% 4
	&+&
	\left(1-\frac{\xzt^2}{4}\right)^{-2} 
	\log \left(\xzt^2\right) 
		\left[ 
		 \underbrace{\left(\frac{16 \xwt^2 \xwz^2}{9}-\frac{31 \xwt^2}{18}+\frac{371 \xzt^2}{288}\right)}_{-0.420184}
		\right.
	\nonumber\\
	&-&
	 \underbrace{\left(\xwt^4-\xwt^2 \xzt^2+\frac{63 \xzt^4}{64}\right)}_{0.0923025}
	\left.
	+ \underbrace{\left(\frac{85 \xwt^4 \xzt^2}{216}-\frac{199 \xwt^2 \xzt^4}{432}+\frac{319 \xzt^6}{864}\right)}_{-0.00784925}
	\right.
	\nonumber\\
	&-&
		 \underbrace{\left(\frac{4 \xwt^4 \xzt^4}{27}-\frac{5 \xwt^2 \xzt^6}{27}+\frac{163 \xzt^8}{1728}\right)}_{0.000348639}
	\left.
			+ \underbrace{\left(\frac{\xwt^4 \xzt^6}{54}-\frac{5 \xwt^2 \xzt^8}{216}+\frac{17 \xzt^{10}}{1728}\right)}_{-7.3468\times 10^{-6}} 
	\right]
	\nonumber\\
% 5
	&+&
	\underbrace{\left(1-\frac{\xht^2}{4}\right)^{-1}\left(-\frac{\xht^6}{48}+\frac{5 \xht^4}{24}-\frac{9 \xht^2}{16}-\frac{1}{8}\right)  \log \left(\xht^2\right)}_{0.270639}
	\nonumber\\
% 6 + 7
	&-& 
	\underbrace{\frac{1- \xht^2}{8\xht^2} %\left(\frac{1}{8}-\frac{1}{8 \xht^2}\right) 
	\left(1-\frac{\xht^2}{4}\right)^{-1} \Phi\left(\frac{\xht^2}{4}\right)}_{-0.19534}
+	\left(1-\frac{\xzt^2}{4}\right)^{-2} \Phi\left(\frac{\xzt^2}{4}\right) 
	\times
	\nonumber\\
& \times & 
	\left[ 
		\underbrace{-\frac{2 \xwz^4}{3}+\frac{5 \xwz^2}{6}-\frac{7}{96}}_{0.188693}
	+		\underbrace{\left(\frac{\xwt^2 \xwz^2}{3}-\frac{5 \xwt^2}{12}+\frac{\xzt^2}{12}\right)}_{-0.0119025}
	\right]
	\nonumber\\
	&+ &
	 \sqrt{\left(\frac{4}{\xzt^2} -1\right)}
	\arccos\left(\frac{\xzt}{2}\right)
	\left[ 
	\underbrace{\left(\frac{32 \xwt^4}{27}-\frac{40 \xwt^2 \xzt^2}{27}+\frac{7 \xzt^4}{54}\right)}_{-0.112343}
	\right.
\nonumber\\
	&+&
	\left.
		\underbrace{\left(\frac{16 \xwt^4 \xzt^2}{27}-\frac{20 \xwt^2 \xzt^4}{27}+\frac{17 \xzt^6}{54}\right)}_{0.00987727} 
	\right]
	-
	\underbrace{\frac{\xht^2}{6}}_{-0.0877755}
		\nonumber\\
	&- &
	     \frac{\left(1-\xwt^2\right)^2}{3} 
	     \left(\underbrace{\frac{1}{2}}_{0.0248873} + \underbrace{\vphantom{\frac{1}{2}}\xwt^2}_{0.0107204} \right)
	     		\log \left(1-\xwt^2\right),
	\label{eq:as_dec_asal_non_log_contrib} 
\end{eqnarray}
}
where 
\begin{equation}
	\Phi(z) = 4 \sqrt{\frac{z}{1-z}} \mathrm{Cl}_2\left( 2 \arcsin \sqrt z \right)
	\label{eq:phi_func_def}
\end{equation}
	is expressed in terms of the Clausen function of the second order and is coming from  a two-loop massive vacuum integral with two equal masses. 
In order to illustrate how different terms, grouped by the powers of $\xwt \simeq \xzt$, 
	contribute to the {\em final result} for $X^{(0)}_{\as\al}$ and $\XX{1}{\as\al}$, we evaluate them 
for the current PDG \cite{Agashe:2014kda} central values of masses $\mtpole = 173.21(88)$~GeV, $m_W = 80.385(15)$~GeV, $m_Z = 91.1876(21)$~GeV, and $m_h = 125.7(4)$~GeV. 
One can see various cancellations between terms in \eqref{eq:as_dec_asal_contrib_}, rendering\footnote{The indicated uncertainty is associated with
that of input parameters.} $\XX{0}{\as\al} = -1.17(2)$ 
and $\XX{1}{\as\al} = -0.034(15)$. %-0.0348857$ 

It should be mentioned that the scale dependence of the corrections was cross-checked by taking a derivative of \eqref{eq:as_dec} with respect to $\mu$ and 
	comparing the coefficients of the powers of $\as$ and $\al$ appearing in the left- and right-hand side of the equation.
	
Let us estimate an effect due to the neglected $b$-quark Yukawa interactions. 
In the normalization used above the corresponding contribution to $\XX{0}{\as\al}$ and $\XX{1}{\as\al}$ 
should be suppressed by the factor $m_b^2/\mmt\simeq\mathcal{O}(10^{-4})$ giving rise to a number, which is, obviously, much smaller than the leading terms.

In order %to use the decoupling relation \eqref{eq:as_dec} 
to find the value of the strong coupling defined in the SM
from that of effective $\QCD\times\QED$, one should  invert Eq.~\eqref{eq:as_dec} in perturbation theory
\begin{eqnarray}
		\asSM    & = &   \asQCD 
		\left( 1 + \frac{\asQCD}{4\pi} \dzeta{\asQCD}{1}
		+ \frac{\asQCD^2}{(4\pi)^2} \dzeta{\asQCD}{2}
	+ \frac{\asQCD \al}{(4\pi)^2} \dzeta{\asQCD\al }{2}
			\right) 
			\nonumber\\
			\dzeta{\asQCD}{1}  & = &   - \dzeta{\as}{1} = - \frac{2}{3} \lnMTmu 
			\nonumber\\
		\dzeta{\asQCD}{2}  & = &   - \left( \dzeta{\as}{2} - 2 (\dzeta{\as}{1})^2 \right)
		= \frac{14}{3} - \frac{38}{3} \lnMTmu + \frac{4}{9} \lnSqMTmu
			\nonumber\\
			\dzeta{\asQCD\al}{2}  & = &  -\dzeta{\as\al}{2} = -\frac{\mmt}{\mmW \sw^2} \left(\XX{0}{\as\al} + \XX{1}{\as\al} \lnMTmu\right).
\label{eq:as_dec_inverted}
\end{eqnarray}
If we choose the $\mu = M_Z$ scale for matching and use again the PDG world averages to assume that $\asQCD(M_Z) = \as^{(5)}(M_Z) = 0.1185$ and $1/\al(M_Z) = 127.94$,
	equation \eqref{eq:as_dec_inverted} can be casted into
\begin{equation}
       \asSM(M_Z) =  0.1185 \cdot \left[ 1 - \underbrace{0.008067}_{\as} - \underbrace{0.000965}_{\as^2} + \underbrace{0.000143}_{\as\al} 
       + \underbrace{0.000018}_{\as^3}
       \right],        
\label{eq:asMZ_dec_num}
\end{equation}
	in which we also include the three-loop pure QCD contribution from the top quark evaluated by means of the {\tt RunDec} package \cite{Chetyrkin:2000yt}.
One can see that in the considered setup the calculated two-loop contribution has an opposite sign in comparison with the $\mathcal{O}(\as^2)$ 
	part and almost one order of magnitude larger than the three-loop correction due to the strong interactions.

\begin{figure}[th]
\begin{center}
\includegraphics[width=0.7\textwidth]{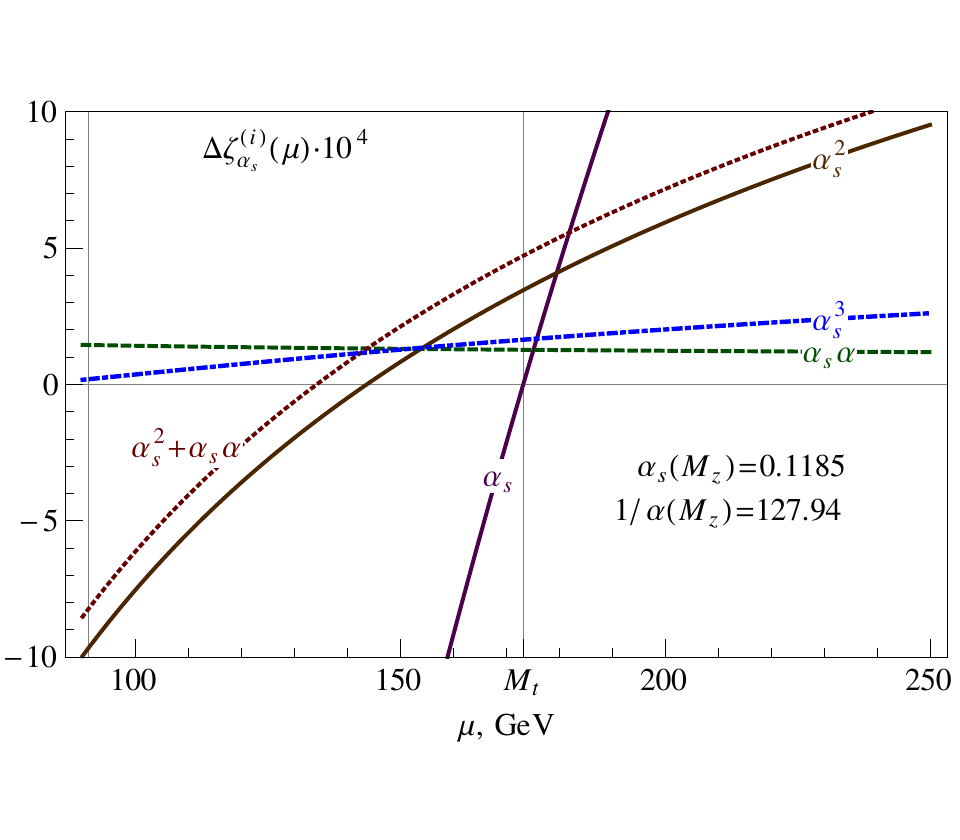}
\end{center}
\caption{Scale dependence of different contributions to the $\as$ matching relation~\eqref{eq:as_dec_inverted}
	around $\mu=M_t$.
	Boundary values for $\as(M_Z)$ and $\al(M_Z)$ are shown. 
	The dependence of $\al$ on $\mu$ is neglected 
	in $(4\pi)^2 \Delta\zeta_{\as}^{(\as\al)}\equiv \asQCD\al\cdot \dzeta{\asQCD\al}{2}$.
\label{fig:dec_scale_dep}
	}
\end{figure}
In order to demonstrate how the value of the calculated correction changes with the decoupling scale $\mu$
we plot the scale dependence of various 
contributions to relation \eqref{eq:as_dec_inverted} (see Fig.~\ref{fig:dec_scale_dep}). 
For convenience, we introduce some obvious notation
\begin{equation}
  \Delta \zeta^{(\as)}_{\as} \equiv \frac{\asQCD}{(4\pi)} \dzeta{\asQCD}{1},~
  \Delta \zeta^{(\as^2)}_{\as} \equiv \frac{\asQCD^2}{(4\pi)^2} \dzeta{\asQCD}{2},~\mathrm{etc.}
\end{equation}
	The RG evolution of $\asQCD(\mu)$ is found by means of four-loop RGEs for $n_f=5$ QCD
	and the scale dependence of the fine-structure constant is neglected. 
From Fig.~\ref{fig:dec_scale_dep} one can deduce that the 
	electroweak correction is at the per myriad level and tends to decrease with the matching scale while 
	pure QCD contributions become larger with $\mu$.
At the scale $\mu=\mu_0\simeq 155$~GeV they meet each other and for $\mu\gtrsim\mu_0$ 
	the three-loop $\mathcal{O}(\as^3)$  contribution becomes larger than the $\mathcal{O}(\as\al)$ terms. 
Moreover, it is interesting to note that at the same scale also 
	the $\mathcal{O}(\as^2)$ line crosses the meeting point.
As a consequence, addition of the electroweak correction to the well-known $\mathcal{O}(\as^2)$ 
	expression doubles the two-loop QCD result. 

One can also notice that the two-loop $\mathcal{O}(\as\al)$ contributions to the decoupling of $\as$ 
	at certain values of the matching scale can compete with the pure QCD corrections usually included
	in the RG analysis of the Standard Model.
Due to this, for a precision study of the latter one has to take the calculated correction into account.
However, one should keep in mind that to avoid double-counting it should be included 
	only if the input value $\asQCD(\mu \sim M_Z)$ is obtained from some low-energy observable.

It is also worth mentioning that recent re-analysis of the ALEPH data for hadronic $\tau$-decays 
	\cite{Boito:2014sta} yields a value for $\as(m_\tau)=0.303(14)$, which is lower 
	than the pre-averaged result $\as(m_\tau) = 0.330(14)$ quoted in PDG~\cite{Agashe:2014kda}.
When evolved to the reference scale and matched with the SM by taking into account only QCD interactions
	of the $c-$, $b-$, and $t$-quarks it leads to a bit lower value for the SM strong coupling,
	which, in turn, could be accounted for (at least partially) by the inclusion of 
	the electroweak contribution presented here.

To conclude, the two-loop $\mathcal{O}(\as\al)$ matching corrections for the strong coupling 
are calculated\footnote{The corresponding expressions in a computer-readable form 
	can be found online as ancillary files of the arXiv version of the paper.} and analyzed numerically. 
It is found that for the matching scale varying from 100 to 250 GeV the corrections tend to ``screen'' the
	SM strong coupling, while the pure one- and two-loop QCD contributions lead to additional ``anti-screening''
	up to $\mu\lesssim 150-175$ GeV.

The corrections can play a role in reducing theoretical uncertainties of the running \MSbar-coupling $\as$
	at the electroweak scale so that the SM can be tested in both the low-energy and high-energy 
	region with increased accuracy.

\subsection*{Acknowledgments}
We are grateful to A.V.~Nesterenko, M.~Kalmykov, O.~Veretin, A.~Pikelner, and B.Kniehl 
for fruitful discussions.
We also thank S.~Martin for pointing out a small typo in \eqref{eq:as_dec_asal_non_log_contrib} and confirming our results in Ref.~\cite{Martin:2018yow}.  
Furthermore, the author would like to thank 2nd Institute for Theoretical Physics of the University of Hamburg
	for hospitality while this article was being finished. 
This work is supported in part by RFBR 
	grant 12-02-00412-a, 
	the Ministry of Education and Science of the Russian Federation, Grant MK-1001.2014.2,
	Heisenberg-Landau Programme, 
	and
	by the German Research Foundation through
	the  Collaborative Research Center No.~676 {\it Particles, Strings and the
	Early Universe---The Structure of Matter and Space Time}.


\begin{thebibliography}{10}

\bibitem{Moch:2014tta}
S. Moch et~al.,
\newblock (2014), 1405.4781.

\bibitem{'tHooft:1973mm}
G. 't~Hooft,
\newblock Nucl.Phys. B61 (1973) 455.

\bibitem{vanRitbergen:1997va}
T. van Ritbergen, J. Vermaseren and S. Larin,
\newblock Phys.Lett. B400 (1997) 379, hep-ph/9701390.

\bibitem{Czakon:2004bu}
M. Czakon,
\newblock Nucl.Phys. B710 (2005) 485, hep-ph/0411261.

\bibitem{Bernreuther:1981sg}
W. Bernreuther and W. Wetzel,
\newblock Nucl.Phys. B197 (1982) 228.

\bibitem{Steinhauser:2002rq}
M. Steinhauser,
\newblock Phys.Rept. 364 (2002) 247, hep-ph/0201075.

\bibitem{Grozin:2012ec}
A. Grozin,
\newblock Int.J.Mod.Phys. A28 (2013) 1350015, 1212.5144.

\bibitem{Chetyrkin:1997un}
K. Chetyrkin, B.A. Kniehl and M. Steinhauser,
\newblock Nucl.Phys. B510 (1998) 61, hep-ph/9708255.

\bibitem{Chetyrkin:2005ia}
K. Chetyrkin, J.H. Kuhn and C. Sturm,
\newblock Nucl.Phys. B744 (2006) 121, hep-ph/0512060.

\bibitem{Schroder:2005hy}
Y. Schroder and M. Steinhauser,
\newblock JHEP 0601 (2006) 051, hep-ph/0512058.

\bibitem{Kniehl:2006bg}
B. Kniehl et~al.,
\newblock Phys.Rev.Lett. 97 (2006) 042001, hep-ph/0607202.

\bibitem{Grozin:2011nk}
A.G. Grozin et~al.,
\newblock JHEP 1109 (2011) 066, 1107.5970.

\bibitem{Chetyrkin:2000yt}
K. Chetyrkin, J.H. Kuhn and M. Steinhauser,
\newblock Comput.Phys.Commun. 133 (2000) 43, hep-ph/0004189.

\bibitem{Jegerlehner:2012kn}
F. Jegerlehner, M.Y. Kalmykov and B.A. Kniehl,
\newblock Phys.Lett. B722 (2013) 123, 1212.4319.

\bibitem{Kniehl:2014yia}
B.A. Kniehl and O.L. Veretin,
\newblock Nucl.Phys. B885 (2014) 459, 1401.1844.

\bibitem{Degrassi:2012ry}
G. Degrassi et~al.,
\newblock JHEP 1208 (2012) 098, 1205.6497.

\bibitem{Bezrukov:2012sa}
F. Bezrukov et~al.,
\newblock JHEP 1210 (2012) 140, 1205.2893.

\bibitem{Luo:2002ey}
M.x. Luo and Y. Xiao,
\newblock Phys.Rev.Lett. 90 (2003) 011601, hep-ph/0207271.

\bibitem{Mihaila:2012fm}
L.N. Mihaila, J. Salomon and M. Steinhauser,
\newblock Phys.Rev.Lett. 108 (2012) 151602, 1201.5868.

\bibitem{Chetyrkin:2012rz}
K. Chetyrkin and M. Zoller,
\newblock JHEP 1206 (2012) 033, 1205.2892.

\bibitem{Bednyakov:2012rb}
A. Bednyakov, A. Pikelner and V. Velizhanin,
\newblock JHEP 1301 (2013) 017, 1210.6873.

\bibitem{Chetyrkin:2013wya}
K. Chetyrkin and M. Zoller,
\newblock JHEP 1304 (2013) 091, 1303.2890.

\bibitem{Bednyakov:2013eba}
A. Bednyakov, A. Pikelner and V. Velizhanin,
\newblock Nucl.Phys. B875 (2013) 552, 1303.4364.

\bibitem{Bednyakov:2013cpa}
A. Bednyakov, A. Pikelner and V. Velizhanin,
\newblock Nucl.Phys. B879 (2014) 256, 1310.3806.

\bibitem{Bednyakov:2014pia}
A. Bednyakov, A. Pikelner and V. Velizhanin,
\newblock Phys.Lett. B737 (2014) 129, 1406.7171.

\bibitem{Harlander:2005wm}
R. Harlander, L. Mihaila and M. Steinhauser,
\newblock Phys.Rev. D72 (2005) 095009, hep-ph/0509048.

\bibitem{Bednyakov:2007vm}
A. Bednyakov,
\newblock Int.J.Mod.Phys. A22 (2007) 5245, 0707.0650.

\bibitem{Bauer:2008bj}
A. Bauer, L. Mihaila and J. Salomon,
\newblock JHEP 0902 (2009) 037, 0810.5101.

\bibitem{Kurz:2012ff}
A. Kurz, M. Steinhauser and N. Zerf,
\newblock JHEP 1207 (2012) 138, 1206.6675.

\bibitem{Bednyakov:2009wt}
A. Bednyakov,
\newblock Int.J.Mod.Phys. A25 (2010) 2437, 0912.4652.

\bibitem{Noth:2010jy}
D. Noth and M. Spira,
\newblock JHEP 1106 (2011) 084, 1001.1935.

\bibitem{Agashe:2014kda}
Particle Data Group, K. Olive et~al.,
\newblock Chin.Phys. C38 (2014) 090001.

\bibitem{Pich:1998xt}
A. Pich,
\newblock (1998) 949, hep-ph/9806303.

\bibitem{Larin:1994va}
S. Larin, T. van Ritbergen and J. Vermaseren,
\newblock Nucl.Phys. B438 (1995) 278, hep-ph/9411260.

\bibitem{Fleming:2007qr}
S. Fleming et~al.,
\newblock Phys.Rev. D77 (2008) 074010, hep-ph/0703207.

\bibitem{Fleischer:1980ub}
J. Fleischer and F. Jegerlehner,
\newblock Phys.Rev. D23 (1981) 2001.

\bibitem{Awramik:2002vu}
M. Awramik et~al.,
\newblock Phys.Rev. D68 (2003) 053004, hep-ph/0209084.

\bibitem{Hempfling:1994ar}
R. Hempfling and B.A. Kniehl,
\newblock Phys.Rev. D51 (1995) 1386, hep-ph/9408313.

\bibitem{Jegerlehner:2002em}
F. Jegerlehner, M.Y. Kalmykov and O. Veretin,
\newblock Nucl.Phys. B658 (2003) 49, hep-ph/0212319.

\bibitem{Surguladze:1996hx}
L.R. Surguladze,
\newblock (1996), hep-ph/9803211.

\bibitem{Gross:1973id}
D.J. Gross and F. Wilczek,
\newblock Phys.Rev.Lett. 30 (1973) 1343.

\bibitem{Politzer:1973fx}
H.D. Politzer,
\newblock Phys.Rev.Lett. 30 (1973) 1346.

\bibitem{Caswell:1974gg}
W.E. Caswell,
\newblock Phys.Rev.Lett. 33 (1974) 244.

\bibitem{Jones:1974mm}
D. Jones,
\newblock Nucl.Phys. B75 (1974) 531.

\bibitem{Tarasov:1976ef}
O. Tarasov and A. Vladimirov,
\newblock Sov.J.Nucl.Phys. 25 (1977) 585.

\bibitem{Egorian:1978zx}
E. Egorian and O. Tarasov,
\newblock Teor.Mat.Fiz. 41 (1979) 26.

\bibitem{Hahn:2000kx}
T. Hahn,
\newblock Comput.Phys.Commun. 140 (2001) 418, hep-ph/0012260.

\bibitem{Davydychev:1992mt}
A.I. Davydychev and J. Tausk,
\newblock Nucl.Phys. B397 (1993) 123.

\bibitem{Chetyrkin:1981qh}
K. Chetyrkin and F. Tkachov,
\newblock Nucl.Phys. B192 (1981) 159.

\bibitem{Boito:2014sta}
D. Boito et~al.,
\newblock (2014), 1410.3528.

\bibitem{Martin:2018yow}
S.P. Martin,
\newblock (2018), 1812.04100.

\end{thebibliography}
\end{document}